\documentclass[10pt, paper=a4, UKenglish]{article}
\usepackage{graphicx}
\usepackage{amsmath}
\def\Title#1{\begin{center} {\Large #1 } \end{center}}
\def\Author#1{\begin{center}{ \sc #1} \end{center}}
\def\Address#1{\begin{center}{ \it #1} \end{center}}

\newcommand\pubblock{\rightline{\begin{tabular}{l} Proceedings of the CTD 2023\\ \pubnumber\\
         \pubdate  \end{tabular}}}

\newenvironment{Abstract}{\begin{quotation} \begin{center}
             \large ABSTRACT \end{center}\bigskip
      \begin{center}\begin{large}}{\end{large}\end{center} \end{quotation}}

\newenvironment{Presented}{\begin{quotation} \begin{center}
             PRESENTED AT\end{center}\bigskip
      \begin{center}\begin{large}}{\end{large}\end{center} \end{quotation}}

\def\Acknowledgements{\bigskip  \bigskip \begin{center} \begin{large}
      \bf ACKNOWLEDGEMENTS \end{large}\end{center}}

\def\beq{\begin{equation}}
\def\eeq#1{\label{#1}\end{equation}}
\def\eeqn{\end{equation}}

\def\beqa{\begin{eqnarray}}
\def\eeqa#1{\label{#1}\end{eqnarray}}
\def\eeqan{\end{eqnarray}}

\let\bar=\overbar

\def\Dslash{\not{\hbox{\kern-4pt $D$}}}
\def\dslash{\not{\hbox{\kern-2pt $\del$}}}

\def\msb{{\bar{\ssstyle M \kern -1pt S}}}

\usepackage{color}
\usepackage{lineno}
\usepackage{hyperref}
\usepackage[capitalise]{cleveref}
\usepackage{siunitx}
\usepackage{caption}
\usepackage{subcaption}
\usepackage{adjustbox}

\usepackage[style=numeric,backend=biber,natbib=true]{biblatex}
\newcommand\pubnumber{PROC-CTD2023-27}

\newcommand\pubdate{\today}

\newcommand{\conference}{Connecting the Dots Workshop (CTD 2023)\\
October 10-13, 2023}

\usepackage{fancyhdr}
\definecolor{mygrey}{RGB}{105,105,105}
\begin{document}


\large
\begin{titlepage}
\pubblock

\vfill
\Title{Combined track finding with GNN \& CKF}
\vfill

\Author{Lukas Heinrich\textsuperscript{1}, Benjamin Huth\textsuperscript{2}, Andreas Salzburger\textsuperscript{3}, Tilo Wettig\textsuperscript{2}}
\Address{\textsuperscript{1}Technical University of Munich, \textsuperscript{2}University of Regensburg, \textsuperscript{3}CERN}
\vfill

\begin{Abstract}
The application of Graph Neural Networks (GNN) in track reconstruction is a promising approach to cope with the challenges arising at the High-Luminosity upgrade of the Large Hadron Collider (HL-LHC). GNNs show good track-finding performance in high-multiplicity scenarios and are naturally parallelizable on heterogeneous compute architectures.

Typical high-energy-physics detectors have high resolution in the innermost layers to support vertex reconstruction but lower resolution in the outer parts. GNNs mainly rely on 3D space-point information, which can cause reduced track-finding performance in the outer regions.

In this contribution, we present a novel combination of GNN-based track finding with the classical Combinatorial Kalman Filter (CKF) algorithm to circumvent this issue: The GNN resolves the track candidates in the inner pixel region, where 3D space points can represent measurements very well. These candidates are then picked up by the CKF in the outer regions, where the CKF performs well even for 1D measurements.

Using the ACTS infrastructure, we present a proof of concept based on truth tracking in the pixels as well as a dedicated GNN pipeline trained on $t\bar{t}$ events with pile-up 200 in the OpenDataDetector.
\end{Abstract}

\vfill

\begin{Presented}
\conference
\end{Presented}
\vfill
\end{titlepage}
\setcounter{footnote}{0}
%

\normalsize



\section{Introduction and motivation}
\label{sec:intro}

The upcoming upgrade of the \emph{Large Hadron Collider} (LHC) towards higher luminosities (HL-LHC) is a major challenge for track-reconstruction software. It is expected that the number of simultaneous proton-proton collisions (pile-up) will increase from currently $\left\langle \mu \right\rangle = 50 $ in LHC Run 3 up to $\left\langle \mu \right\rangle = 200$, which will bring the current CPU-based reconstruction software to its limit and beyond. For that reason, much effort is put into exploring new algorithms, especially machine-learning algorithms based on neural networks because these are parallelizable very efficiently on heterogeneous architectures like GPUs.

Recently, \emph{Graph Neural Network} (GNN)-based algorithms have shown very promising results for track finding~\cite{exatrkx, gnn4itk}. However, in their basic version, their performance suffers from low spatial resolution in detector regions like strip modules. 

Proposed solutions to this problem include more complex network architectures like heterogeneous GNNs~\cite{hetero_gnn}. However, this work explores a different approach: The idea is to use the GNN for track finding in the inner, high-resolution parts of the detector, where it was shown to be able to resolve the track candidates very efficiently \cite{Calafiura:2022zft}. For the outer parts, we rely on a standard algorithm for track finding, the \emph{Combinatorial Kalman Filter} (CKF),\footnote{The CKF is a combined track-finding and track-fitting algorithm that combinatorially explores track candidates by considering multiple measurements per detector module.} for which the GNN track candidates serve as seeds. For the track fitting, we use the CKF both in the inner and the outer parts. 
From combining GNN and CKF, we expect several benefits:

\begin{itemize}
	\item \textit{Improved seed quality}. Triplet-seeding algorithms usually produce many duplicate seeds that result in duplicate tracks. GNN seeds can be larger than 3 hits and are unlikely to produce duplicate seeds.\footnote{In the case of no merged-hit clusters, duplicate seeds with shared hits are not allowed by construction.} This should improve the computational performance and avoid track cleaning by an ambiguity resolution. afterward.
	\item \textit{Less branching}. By construction, we do not allow the CKF to branch in the pixel region, where it follows the track candidate predicted by the GNN. We also expect that due to high-quality seeds, we can reduce the branching in the outer regions. This will have similar effects on duplication rate and performance as the previous benefit.
	\item \textit{Natural handling of 1D measurements}. The CKF has a natural, mathematical consistent way of handling 1D measurements, whereas GNN architectures, at least currently, rely on 3D space points that must be constructed from measurements and the detector geometry. For a precise 3D-spacepoint construction, a 2D measurement is necessary.
	\item \textit{Smaller graph construction}. Graph construction is a critical part of a GNN pipeline and often expensive in both runtime and memory. By restricting the graph to the pixels, these costs are reduced naturally.
\end{itemize}

Tests of this idea showed room for improvement of the track-finding performance in some regions. Such improvement was achieved by also including hits from the short-strip-barrel section in the GNN.


\section{Experimental setup}
\label{sec:setup}

For the present study, a track reconstruction chain that includes a GNN track-finding module for the pixels and a subsequent CKF that is seeded with the GNN track candidates has been implemented using the ACTS toolkit~\cite{acts}. In the following, we refer to this chain as the GNN+CKF chain. For the detector geometry, the OpenDataDetector (ODD) is used~\cite{odd}, a virtual, hermetic detector with 3 parts: inner, high-resolution pixel modules, a short-strip section, and double-sided long-strip modules in the outermost part.

In addition to this novel tracking chain, several other chains were studied for comparison:
\begin{itemize}
	\item The \emph{proof-of-concept} chain is very similar to the GNN+CKF chain, but the GNN is replaced by a truth-track-finder. This is helpful for figuring out whether inefficiencies come from the GNN or the CKF part of the main chain.
	\item A \emph{standard CKF} chain with classical triplet seeding allows for comparison to state-of-the-art tracking tools. However, we did not tune this chain in this study, and thus we do not expect cutting-edge performance.
	\item As an upper performance bound, a \emph{Kalman truth-tracking} chain has been implemented as well.
\end{itemize}

For this study, $t\bar{t}$-events with pile-up 200 were generated using the Pythia8~\cite{pythia} event generator, followed by a Geant4~\cite{geant4} full simulation in the ODD. To emulate the detector readout, a hybrid approach was chosen:
\begin{itemize}
    \item For the region covered by the GNN, a \emph{geometric digitization} was applied. This digitization implements a basic simulation of the individual cell activation based on the path of the particle through the module, and thus can provide cluster information in addition to the hit coordinates. The only major simplification in this setup is that cluster merging was disabled. This means that each cluster contains only one hit and no shared hits are possible. This simplifies the mapping of a collision event to a graph.
    \item For the remaining regions, a simple digitization based on Gaussian smearing of the truth hits was used, both because the CKF does not directly use cluster information and to limit the effort for validating the digitization configuration.
\end{itemize}

We implemented a GNN-based track-finding pipeline as follows: The first element is a \emph{graph-construction} stage based on metric learning. This is followed by three \emph{edge-classification} stages (one simple filter based on a \emph{multi-layer perceptron} (MLP) and two GNN stages\footnote{It turned out that the second GNN stage improved the performance. We suspect that the improvement is due to a more balanced graph in the second GNN stage. It would be interesting to explore this issue in future work.}). The final element is a \emph{track-building} stage, in which track candidates are constructed from the GNN output by a simple connected-components algorithm. 

The two GNN stages use a modified version of the \emph{Interaction GNN} that is provided in the GNN4ITk Common Framework~\cite{commonframework}. The major differences are the implementation of the undirected graph and the recurrent implementation of the message-passing step.


\section{Training}
\label{sec:training}

The training is done with a custom branch of the GNN4ITk Common Framework, which provides the infrastructure to train and evaluate multi-stage pipelines based on PyTorch~\cite{pytorch}.

For the training of the GNN module, 2000 events with the configuration described above were simulated, of which 250 events each were dedicated for testing and validation purposes, respectively. The 3D space points in cylindrical coordinates $(r, \varphi, z)$ are used as input features to the neural networks. We also used additional features based on the cluster shape (\emph{cell count}, \emph{sum of activations} and the \emph{cluster size} in the surface-local coordinates) in the graph-construction stage because they lead to performance improvements in this stage. All input features are scaled such that they are in the range between 0 and 1. Particle trajectories that loop inside the detector volume were split at the turning points before generating the ground truth so that the network does not have to learn signatures of looping particles.

In order to focus the training on particles of interest, edges of particles with $p_T > \SI{1}{\giga\electronvolt}$ and at least 3 hits in the pixels were assigned higher weight. We call this \emph{target selection} in the following.

The classifier stages each return a score between 0 and 1 for every edge. In the different stages, different score cuts were used to classify the edges into positive and negative edges. For the MLP filter and the first GNN, we have chosen a very low score cut to preserve a very high efficiency and only remove edges that are classified as negative with high confidence. In the last stage, a balanced score cut of $0.5$ was used.

To quantify the performance of different pipeline stages, we define metrics based on classified edges,

\begin{equation}
\text{ edge efficiency} = \frac{\text{true positive edges}}{\text{true edges}}\,, \qquad \text{edge purity} = \frac{\text{true positive edges}}{\text{positive edges}}\,.
\end{equation}

The resulting graph is perfect if both metrics are equal to $1$. We can evaluate these metrics for both the whole set of edges and the target edges. Because we do not expect the pipeline to be a perfect classifier between target and non-target particles, both \emph{target efficiency} and \emph{overall purity} should be close to 1.

%

\begin{figure}[t]
	\centering
	\includegraphics[width=0.99\linewidth]{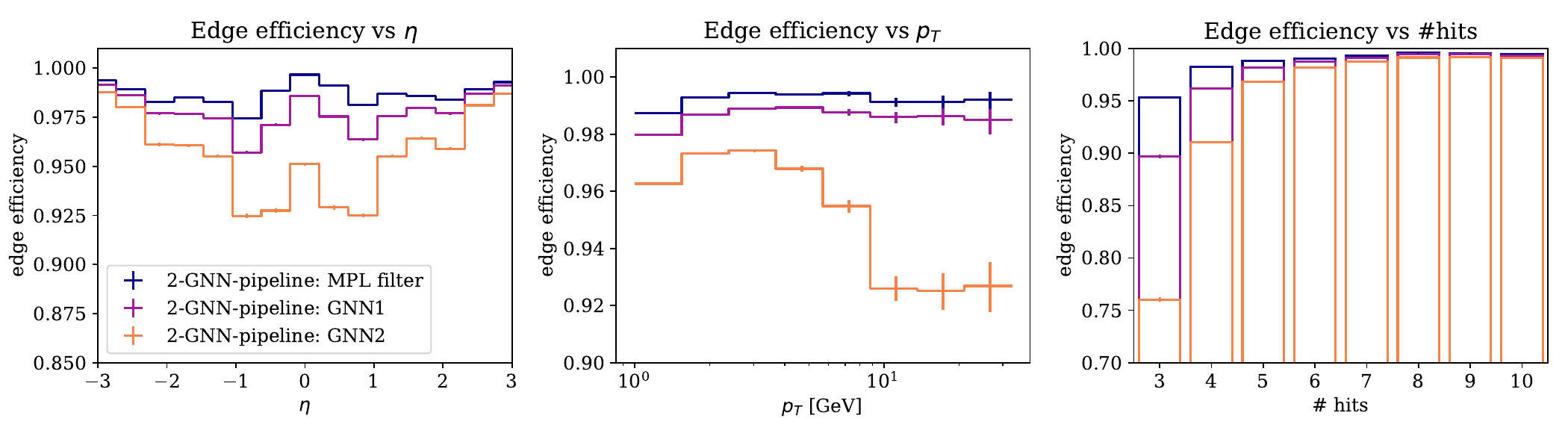}
	\caption{Target-edge efficiency of the different stages as a function of $\eta$, $p_T$ and the number of pixel hits of the corresponding particle for the pipeline trained with the pixel hits only}
	\label{fig:2gnn_edge_eff}
\end{figure}

We reach an overall purity of $>0.99$ at the second GNN, which is close to optimal. Therefore, we focus on the target-edge efficiency in the following (see~\cref{fig:2gnn_edge_eff}).

We observe a very high and mostly flat performance for the first two classifier stages. However, at the second GNN, the performance drops significantly for $|\eta| \approx 1$ and $p_T > \SI{10}{\giga\electronvolt}$. A possible explanation  is the number of hits available to the GNN: The ODD has only 4 pixel layers. For 3 hits in roughly the same $\eta$-slice, each combination of hits is physically valid. Correspondingly, we only observe an efficiency of $\approx 0.75$ in this case. Even a fourth hit might not provide enough discrimination power at high $p_T$.

\begin{figure}[t]
    \centering
    \includegraphics[width=0.99\linewidth]{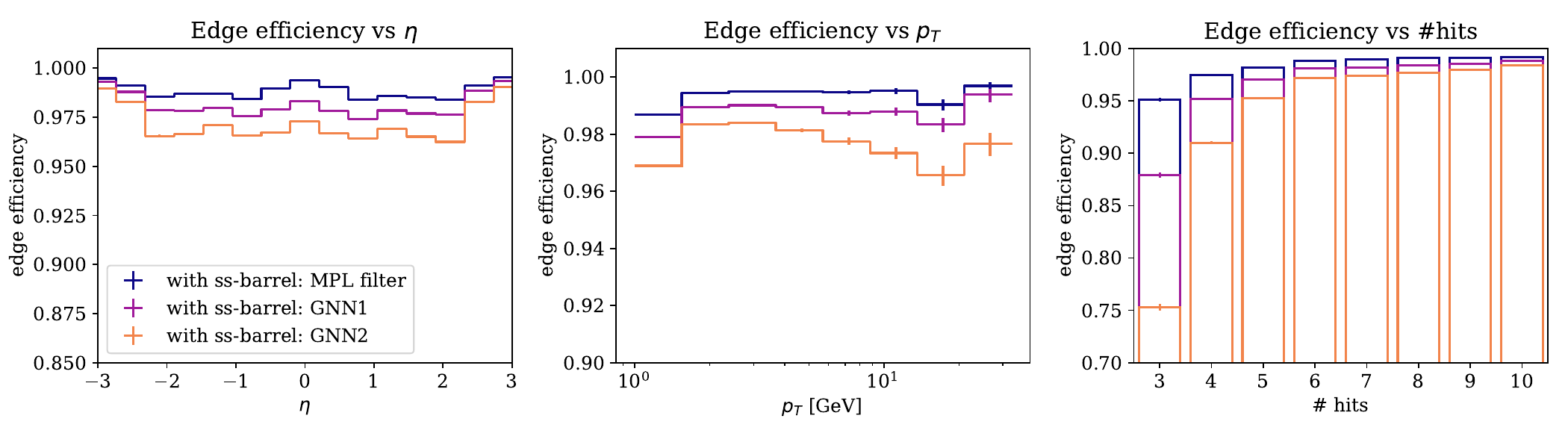}
    \caption{Same as \cref{fig:2gnn_edge_eff} but trained with the pixel hits and the hits in the short-strip barrel.}
    \label{fig:2gnn_edge_eff_with_ssbarrel}
\end{figure}

To provide further evidence for this explanation, we trained the pipeline again, this time also including the short-strip-barrel section (providing 4 extra layers of hits, albeit with greatly reduced resolution). Indeed, we see much better performance (see~\cref{fig:2gnn_edge_eff_with_ssbarrel}). Both the efficiency in the central-barrel region and in the high-$p_T$ region have greatly improved. We will mainly focus on this GNN pipeline in the following.

\section{Inference and performance analysis}
\label{sec:inference}

The implementation of the pipeline in ACTS is based on the C++ interface to PyTorch. A modified CKF algorithm was implemented, which accepts the GNN track candidates with their corresponding start parameters as input. The start parameters are computed from a 3-point estimate from the first, the last and the middle hit of the track candidate. This allows the CKF to only use hits from these track candidates in the pixel region and the short-strip-barrel region. All code to reproduce our results can be found in~\cite{my_repo}.

\begin{figure}[t]
    \centering
    \includegraphics[width=0.9\linewidth]{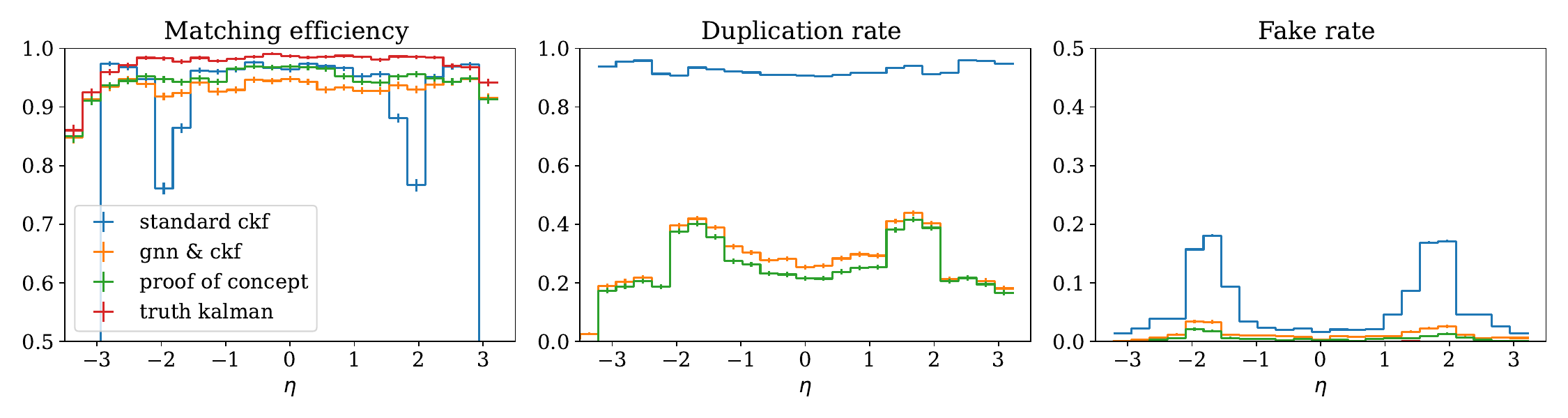}
    \caption{Efficiency, duplication rate, and fake rate vs particle $\eta$ and $p_T$, evaluated with 50 events.}
    \label{fig:track_perf}
\end{figure}

\begin{figure}[t]
    \centering
    \includegraphics[width=0.99\linewidth]{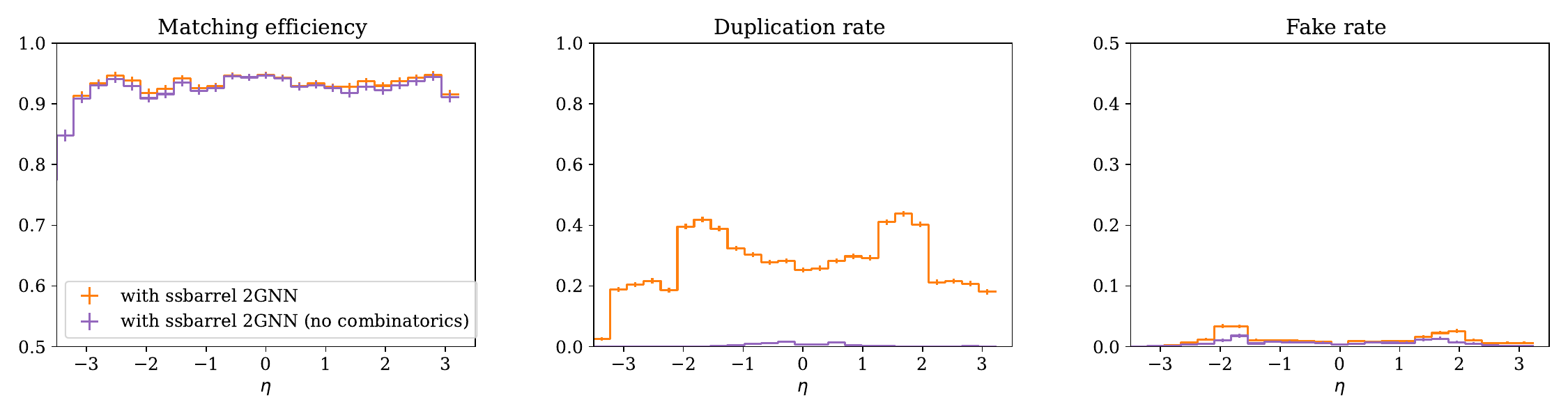}
    \caption{Metrics for the GNN+CKF chain with and without the combinatorial aspect in the CKF stage.}
    \label{fig:remove_c}
\end{figure}

To measure the track-finding performance, we use metrics based on the following selection and matching criteria.
We select particles if they have $p_T > \SI{1}{\giga\electronvolt}$, at least 7 hits in the full detector, and at least 3 hits in the pixels. We select tracks if they have a fitted $p_T > \SI{1}{\giga\electronvolt}$ and at least 7 hits in the full detector. The particles and tracks are matched using \emph{double matching}: We match a track to a particle if the track contains more than \SI{50}{\percent} of the hits of the particle (particle efficiency) and if more than \SI{50}{\percent} of the hits in the track belong to the particle (track purity).  Otherwise, it is a \emph{fake track}.\footnote{The fake track is assigned to its majority particle (based on track purity) for further analysis, see~\cref{fig:track_perf,fig:remove_c}.} Our performance metrics are

\begin{align}
    \text{matching efficiency} &= \frac{N_\text{particles}(\text{selected}, \text{matched})}{N_\text{particles}(\text{selected})} \,, \\
    \text{duplication rate} &= \frac{N_\text{particles}(\text{selected},\text{matched tracks} > 1)}{N_\text{particles}(\text{selected})} \,, \\
    \text{fake rate} &= \frac{N_\text{particles}(\text{selected},\text{unmatched tracks} > 0)}{N_\text{particles}(\text{selected})} \,.
\end{align}
These three metrics measure different aspects of the track-finding performance. In general, we seek high matching efficiency and low fake and duplication rate.

\cref{fig:track_perf} shows the track-finding performance for 50 $t\bar{t}$-events (generated independently of the training set) as histograms of the particle $\eta$. The matching efficiency is shown in~\cref{fig:track_perf} (left). In the central barrel region, we observe a mostly flat performance, with the performance of the standard CKF and the proof-of-concept chain slightly above the GNN+CKF chain. This suggests small inefficiencies in the GNN+CKF chain. For the standard CKF chain, we observe efficiency drops at $|\eta| \approx 2$ and $|\eta| > 3$. This is also reflected in the fake rate and suggests an inefficiency in the classical triplet seeding.

\begin{figure}[t]
    \centering
    \begin{subfigure}[t]{0.47\linewidth}
        \centering
        \adjincludegraphics[width=0.8\linewidth,trim={0 0 {.5\width} 0},clip]{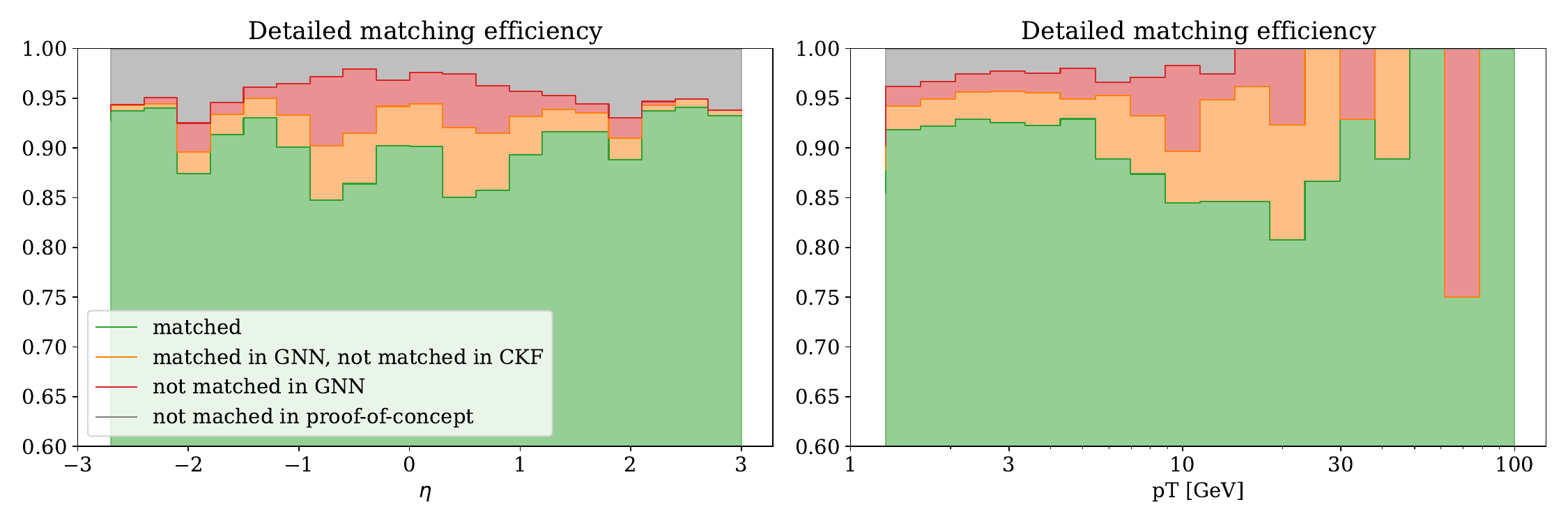}
        \caption{GNN+CKF chain, pixel-only pipeline.}
        \label{fig:detailed_efficiency_pixel_only}
    \end{subfigure}
    \hfill
    \begin{subfigure}[t]{0.47\linewidth}
        \centering
        \adjincludegraphics[width=0.8\linewidth,trim={0 0 {.5\width} 0},clip]{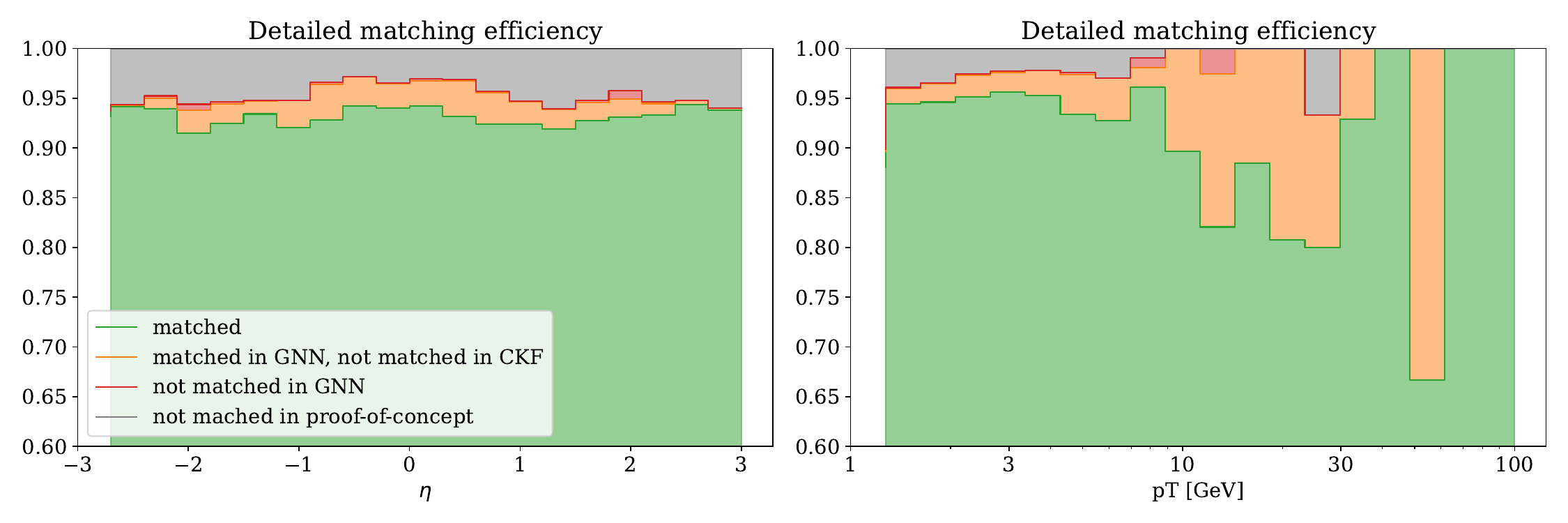}
        \caption{GNN+CKF chain, short-strip-extended pipeline.}
        \label{fig:detailed_efficiency_with_ssstrip}
    \end{subfigure}
    \caption{Detailed matching efficiency for two versions of the GNN+CKF chain, evaluated with 50 events.}
    \label{fig:detailed_efficiency}
\end{figure}

We can resolve the inefficiencies in the GNN+CKF chain in more detail (see~\cref{fig:detailed_efficiency}). We first disregard particles that are not matched in the proof-of-concept chain\footnote{Here the failure to match occurs in the CKF part of the chain because the truth-track finder is perfect by construction.} (shown in gray). Then, we identify particles for which no matching track candidate was found by the GNN (shown in red). This red area is rather large for the pixel-only GNN pipeline (\cref{fig:detailed_efficiency_pixel_only}) but almost disappears for the pipeline that includes the hits in the short-strip barrel (\cref{fig:detailed_efficiency_with_ssstrip}). Finally, some particles with a valid GNN track candidate are not matched after the CKF stage (shown in yellow), mainly because the track candidates are not pure enough to produce high-quality seeds.\footnote{For example, there could be an additional hit included in the track seed that leads to a wrong estimation of the start parameters of the CKF fit.} This could be improved by applying more advanced track-building algorithms.


We return to~\cref{fig:track_perf} and discuss the duplication rate (middle). As expected, the standard CKF has a high duplication rate, which is close to 1 over the whole $\eta$ range. In the proof-of-concept and GNN+CKF chains, duplication is caused mainly by the CKF part. The observation that both chains have similar duplication rates implies that the GNN track candidates, compared to the perfect proof-of-concept candidates, do not lead to a higher duplication rate.

To check if we can reduce this duplication rate further without efficiency costs, we configured the CKF to only accept a single measurement per surface, effectively reducing the CKF to a simple Kalman Filter. The chosen measurement is the one with the lowest $\chi^2$ contribution. The resulting performance is shown in~\cref{fig:remove_c}.
We observe that we can remove the combinatorial aspect of the fitter with negligible matching-efficiency cost. Furthermore, the fake rate is also lower in this setup.


\section{Computational performance}

\begin{figure}[tb]
    \centering
    \begin{subfigure}[t]{0.45\linewidth}
        \centering
        \includegraphics[width=0.8\linewidth]{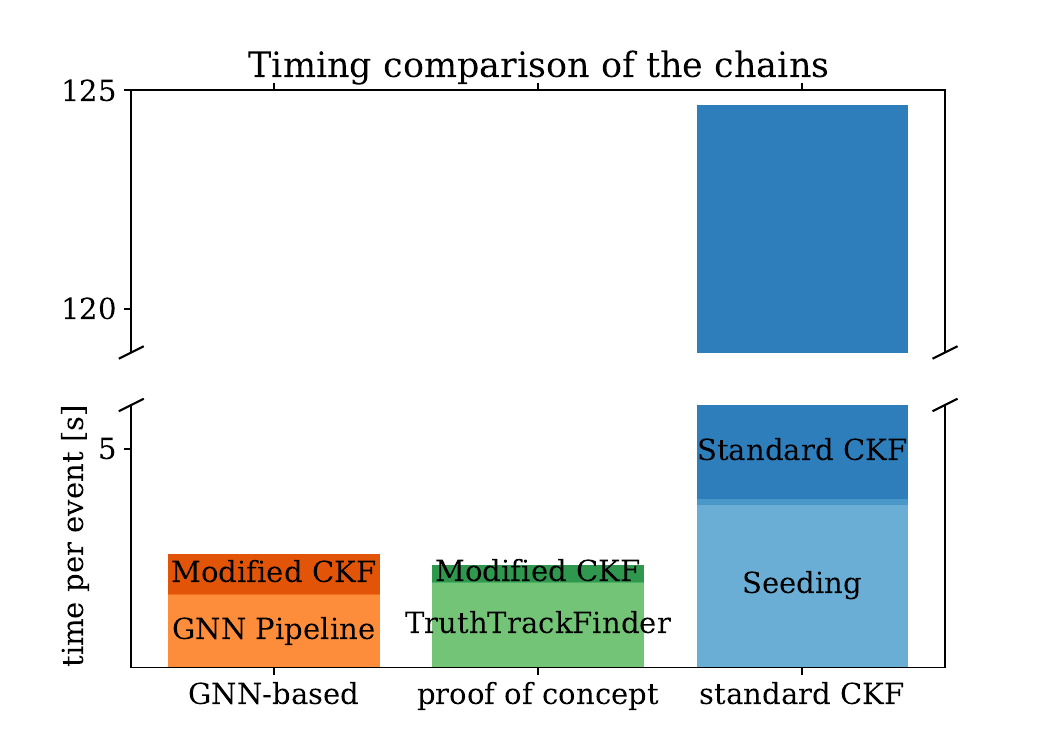} \caption{Timing comparison of different chains.}
    \end{subfigure}
    \hfill
    \begin{subfigure}[t]{0.45\linewidth}
        \centering
        \includegraphics[width=0.8\linewidth]{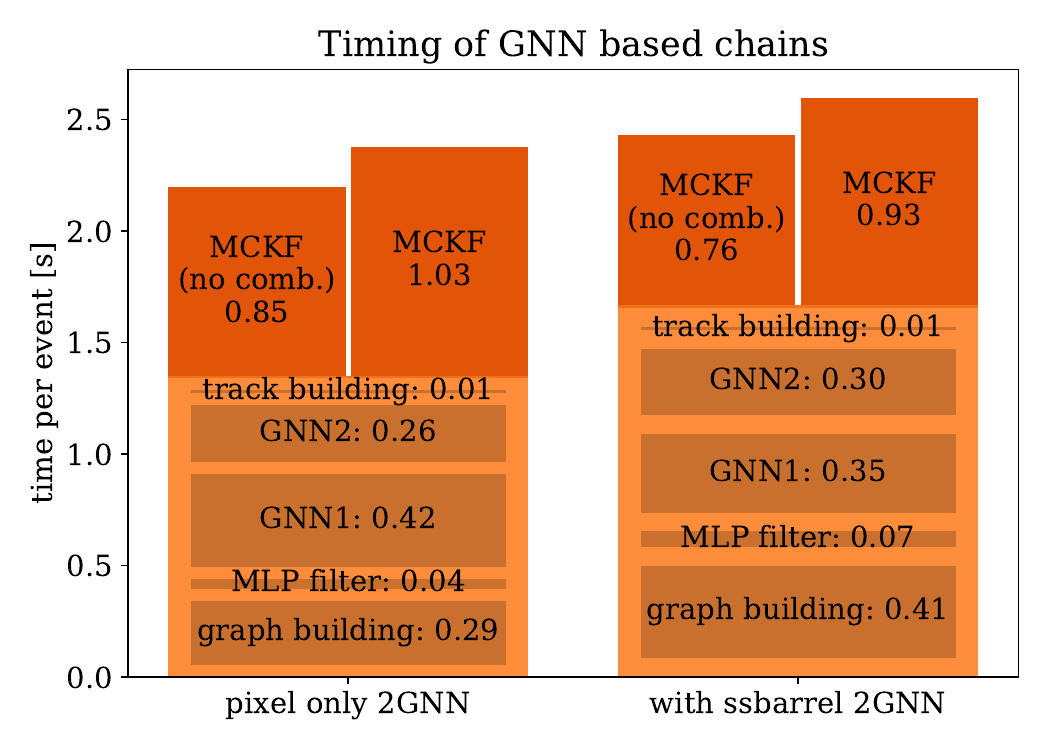}
        \caption{Zoom into the timing of the GNN+CKF chain.}
    \end{subfigure}
    \caption{Average per-event timing for 10 events. MCKF is short for ``Modified CKF''.}
    \label{fig:timing}
\end{figure}

To measure the timing of the different chains, we executed the workflows in a single-threaded environment with a single available GPU. The system comprised an AMD EPYC 7662 64-core processor and an Nvidia A100 with 40GB memory. The results are shown in~\cref{fig:timing}.

The GNN+CKF chain can run the reconstruction of a $t\bar{t}$ event with pile-up 200 in the ODD in $\approx \SI{2}{\second}$. We observe a slightly increased runtime when additional hits from the short-strip barrel are included in the GNN. The standard CKF shows an unreasonably large wall time. This drastic difference in timing is caused by the high duplication rate and reminds us that it is important to keep the combinatorics under control during reconstruction.\footnote{Because the GNN+CKF chain uses the same infrastructure, we can exclude code issues as the problem here.} 


\section{Conclusion}

We have presented a complete track-reconstruction chain that uses a novel hybrid approach by combining a GNN for track finding in the inner detector regions and a subsequent CKF for track finding in the outer parts and fitting.

We observed that using the GNN in the pixels limits its performance in the central detector region. This issue was mitigated by including additional hits from the short-strip-barrel section. This finding suggests that there is a minimum number of hits necessary for the GNN to reach its full discriminative power. It would be interesting to investigate this behavior further in different setups.

We also showed that for good-quality GNN track candidates, the combinatorial aspect in the subsequent CKF can be omitted without significant efficiency loss. As long as Kalman-based fitting is needed, this offers an opportunity to optimize runtime in scenarios with high throughput demands such as triggers.

\Acknowledgements
LH is supported by the Excellence Cluster ORIGINS, funded by the Deutsche Forschungsgemeinschaft (DFG, German Research Foundation) under Germany's Excellence Strategy - EXC-2094-390783311. GPU resources were made available by the ORIGINS Data Science Lab (ODSL). BH and TW are supported in part by the Deutsche Forschungsgemeinschaft (DFG, German Research Foundation), project number 460248186.

\printbibliography


%
%
%
%
%
%
%
%


\end{document}